\documentclass{osa-article}

\journal{oe} 
\articletype{Research Article}

\usepackage{lineno}
\usepackage{tabularx} 
\usepackage{ragged2e} 
\usepackage{booktabs} 
\usepackage{hyperref}
\usepackage{cite}
\usepackage{array}

\hypersetup{
colorlinks=true,
linkcolor=blue,
citecolor=blue,
urlcolor=blue}

\begin{document}

\title{A general design method for ultra-long optical path length multipass matrix cells}

\author{Yiyun Gai,\authormark{1} 
 Wenjin Li,\authormark{1} 
Kaihao Yi,\authormark{1}
Xue Ou,\authormark{1}
Peng Liu,\authormark{1}
 and Xin Zhou\authormark{1,2,*}}

\address{\authormark{1} Center for Advanced Quantum Studies, Applied Optics Beijing Area Major Laboratory, School of Physics and Astronomy, Beijing Normal University, Beijing 100875, China\\
\authormark{2}Key Laboratory of Multiscale Spin Physics (Ministry of Education),Beijing Normal University, Beijing 100875,China\\
}

\email{\authormark{*}zhoux@bnu.edu.cn} 


\begin{abstract*} 
For the first time, we propose a general design method for ultra-long optical path length (OPL) multipass matrix cells (MMCs) based on multi-cycle mode of two-sided field mirrors.
The design idea of the dual circulation mode with two-sided field mirrors is elaborated in detail with the example of MMC based on dual Pickett Bradley White cell (PBWC), and the simple design methods of the other three MMCs based on the dual circulation mode of PBWC and Bernstein Herzberg White cell (BHWC) are given.
Further, we propose a general design method for ultra-long OPL MMCs with multi-cycle mode by adding cyclic elements.
The OPL of the MMCs designed by this method can reach the order of kilometers or even tens of kilometers.
The novel MMCs have the advantages of simple structure, strong spot formation regularity, easy expansion, high mirror utilization ratio, high reuse times of spot spatial position, good stability and extremely high ratio of the  optical path length to the volume (RLV).
In order to evaluate the performance of the new MMCs, an open-path methane gas sensor with the MMC based on triple 
PBWC was constructed, which was used to continuously measure the methane in the laboratory, and the feasibility, effectiveness and practicability of the new design method were verified.
The design method proposed in this paper provides a new idea for the design of multipass cell (MPC), and the new MMCs designed have great potential application value in the field of high-precision trace gas monitoring.
\end{abstract*}

\section{Introduction}
The long-term, continuous and high-precision measurement of the background concentrations of 
major greenhouse gases such as  CH$_4$, CO$_2$ and N$_2$O is of great significance for studying global 
climate change and corresponding greenhouse gas emissions \cite{ref1,ref2,ref3}. High-precision trace gas 
detection is also crucial in the fields of respiratory diagnosis \cite{ref4,ref5}, semiconductor process \cite{ref6} and air pollution control\cite{ref7,ref8,ref9}.
Cavity ring-down spectroscopy (CRDS) \cite{ref10} and off-axis integrated cavity output spectroscopy (OA-ICOS)  \cite{ref11} can meet the stringent measurement accuracy requirements of the above trace gases  \cite{ref12,ref13,ref14} with the help of optical resonators with kilometer-level OPL, but such systems are complex in structure, demanding special measurement environment and high cost. Therefore, it is difficult to achieve large-scale application in complex and diverse practical scenarios.

Tunable diode laser absorption spectroscopy (TDLAS) combined with MPC has the advantages of simple structure, low cost and high reliability, which is the most promising candidate to overcome the above shortcomings  \cite{ref16}.
However, due to the low mirror utilization ratio and the low reuse times of spot spatial position, the OPL of currently available MPCs is limited to the 100-meter level \cite{ref16}, which limits the detection accuracy of the scheme in trace gas measurement.
How to design a MPC with low cost, high stability and long OPL is a hot research topic in recent years, and it is also an urgent challenge to be solved in the application of TDLAS technology in high-precision trace gas detection.

In the past few decades, the most studied and widely used MPCs are the Herriott cell  \cite{ref17,ref18}, the White cell (WC) \cite{ref19}, and their improved versions.
Since elliptical or circular patterns are formed on the Herriott cell mirrors, the utilization rate of the mirrors is low. Some scholars have designed the astigmatic Herriott cell \cite{ref20,ref21,ref22}  by using astigmatic mirrors, which can increase the OPL.
However, the astigmatic mirror is difficult to manufacture, the cost is high, and the spot pattern lacks regularity.
In 2013, Stephen So  \cite{ref23,ref24} proposed two-spherical-mirror-based MPCs with dense patterns, which can form dense spot patterns such as independent circles and concentric circles on the spherical mirrors. Many researchers have carried out the design research of these MPCs and applied them to actual measurement  \cite{ref25,ref29,ref30,ref32,ref33,ref34,WOS:000984854900002}. But the OPL  is usually tens of meters, and there are still problems of uneven spot distribution and low utilization ratio of mirrors.
In 2019, Hudzikowski \cite{ref35} designed a MPC based on multiple spherical mirrors using genetic algorithm, and the design method has also been developed \cite{ref36,ref37,ref38,ref39,ref40}, which improved utilization ratio of mirrors and OPL.
Various improved MPCs based on the Herriott cell reflect only once at the same spatial position, which limits the further improvement of the OPL. The OPL of these types of MPCs is usually within the order of 100 meters.

MMCs based on White cell (WC) are promising solutions to overcome this shortcoming.
Modified versions of the WC, Bernstein Herzberg White cell (BHWC) \cite{ref41} and Pickett Bradley White cell (PBWC) \cite{ref42}, double the OPL by expanding the spot pattern from one row to two rows.
Three-objective Chernin multipass matrix systems (CMMS) \cite{ref43,ref44,ref45,ref46,ref47,ref48} proposed by Chernin formed a parallel combination of multiple PBWCs by adding an objective mirror and a field mirror on the basis of PBWC, and obtained matrix spot pattern on the field mirror.
Four-objective CMMS\cite{ref43,ref44,ref45,ref46,ref49} added two objective mirrors and one field mirror to obtain the combination of PBWCs with spot overlap.
The reuse times of some spot spatial positions on the field mirror is increased to 2.
In the above study, Chernin gave the method of adjusting the number of rows and columns of the matrix, that is, the method of adjusting the OPL.
Guo et al.  \cite{ref49}summarized the general design method of four-objective CMMS.
In order to further improve the utilization ratio of the mirrors, T. Mohamed et al. \cite{ref50} and Xia et al. \cite{ref51} proposed a double-sided matrix spot pattern MMC based on 3 mirrors -3 mirrors, which increases the reuse times of spot spatial position to 4-10, but the regularity of spot pattern is not strong, and the design process is more cumbersome. It needs a computer to run for two days, which is not conducive to its expansion design in application.
At present, there is still a lack of effective design strategies and methods to design MMCs with simple structure, good expansibility, high stability, OPL up to kilometers, which can meet the needs of high-precision detection of trace gases.

Different from Chernin's single-sided field mirror design method, a general design method of ultra-long OPL MMCs based on multi-cycle mode of two-sided field mirrors is proposed in this paper.
Based on the deep study of the reflection law of classical PBWC and BHWC, a design strategy of the dual circulation mode based on two-sided field mirrors is proposed for the first time. Two traditional circular objective mirrors are changed into rectangles, one of which forms new PBWC or BHWC with two spherical mirrors that are added at incident and exit positions of PBWC or BHWC. Four novel MMCs based on PBWC-PBWC, PBWC-BHWC, BHWC-PBWC and BHWC-BHWC are designed, and the simple design method of them is given.

Furthermore, we propose a general design method for ultra-long OPL MMCs with multi-cycle mode by adding cyclic elements, and elaborate the design of MMC based on PBWC-PBWC-PBWC in detail to illustrate the realization of multi-cycle mode. This general design method has a significant effect on the improvement of OPL and RLV.
The novel MMCs have the advantages of simple structure, strong spot formation regularity, easy expansion, high mirror utilization ratio, high reuse times of spot spatial position, good stability and extremely high RLV.

In order to evaluate the performance of the new MMCs, an open-path methane gas sensor with the MMC based on triple PBWC was constructed. The continuous measurement of methane in the laboratory ambient air is realized by using a 1653 nm DFB laser and an InGaAs photodetector, and the feasibility, effectiveness and practicability of the new design method were verified.
The design method proposed in this paper provides a new idea for the design of MPC, and the new MMCs designed have great potential application value in the field of high-precision trace gas monitoring.

\section{ The MMC based on PBWC-PBWC}

PBWC and BHWC are the basic devices of MMCs.
Based on the reflection law of classical PBWC, the design strategy of the dual circulation mode based on two-sided field mirrors is proposed and systematically expounded for the first time in this paper, as well as the detailed configuration and spot pattern of the MMC. A new MMC based on PBWC-PBWC is designed, and its spot pattern formation law is analyzed, the calculation method of the number of ray passes through the MMC is summarized. The important parameters affecting the spot row pitch and column pitch are explored, and the key factors affecting its stability are analyzed. Finally, a simple design method of MMC based on PBWC-PBWC is given.

\subsection{The optical setup of MMC based on PBWC-PBWC}

\begin{figure}[htbp]
\centering\includegraphics[width=10cm]{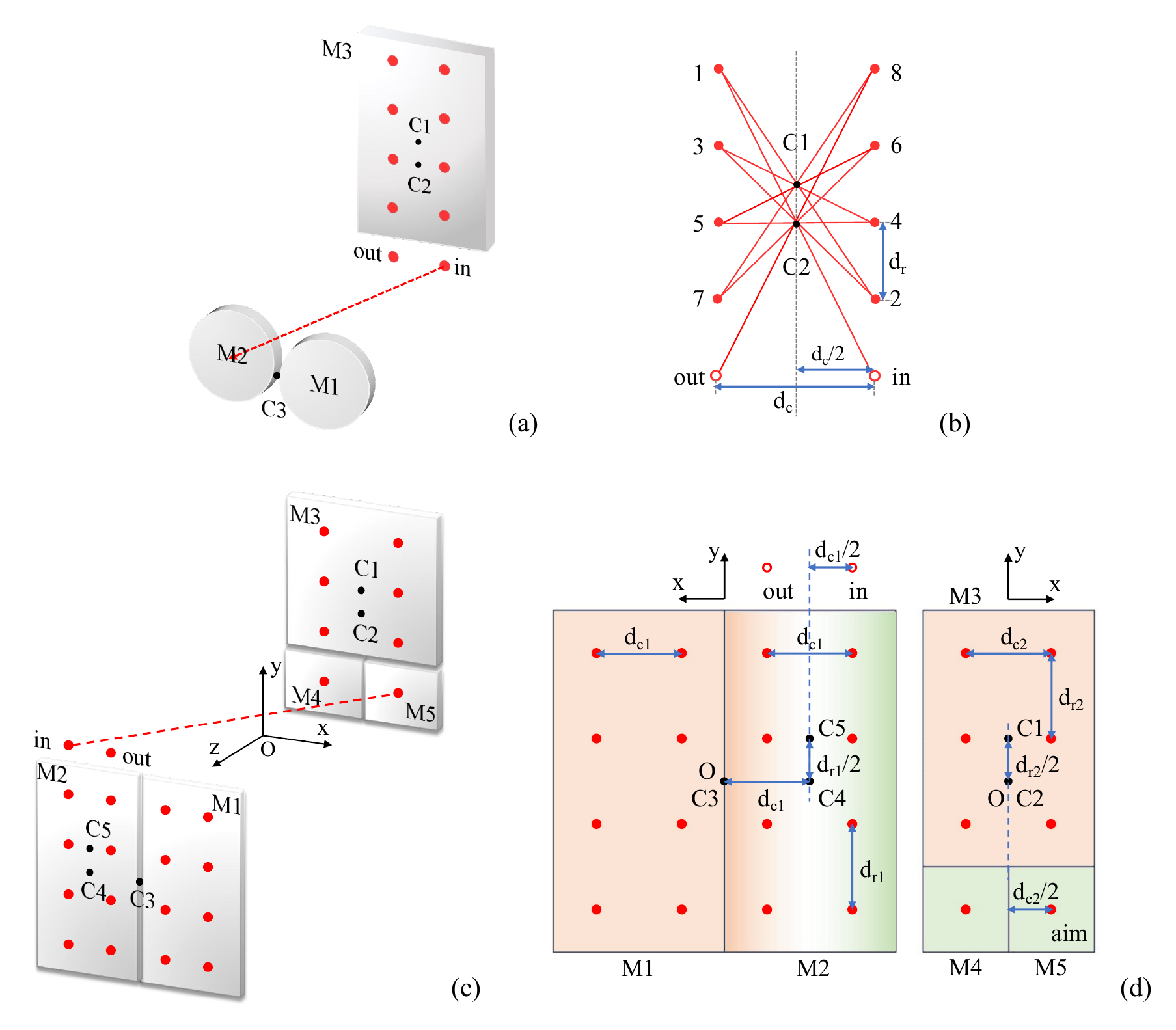}
\caption{Configurations and spot patterns of the classical PBWC and the novel MMC based on PBWC-PBWC. (a) Configuration of the classical PBWC; (b) spot pattern of the field mirror M3 of the classical PBWC; (c) configuration of the novel MMC based on PBWC-PBWC; (d) spot pattern of the novel MMC based on PBWC-PBWC. The O in the center of the mirror is the origin of the coordinate. The x and y directions are shown in the figure, that is, the spot pattern is obtained by standing in the center of the MMC and facing the mirrors. The two-dimensional image of the mirrors in this paper adopts this perspective, which will not be repeated later. The three spherical mirrors marked in green form the PBWC-a, and the three spherical mirrors marked in orange form the PBWC-b.}
\label{fig:1}
\end{figure}

The configuration and spot pattern of the classical PBWC are shown in Fig.\ref{fig:1} (a)(b), which consists of three concave spherical mirrors with equal curvature radius.
For the convenience, we denote the side of the mirrors with incident ray as the incident side, and the other side of the mirrors as the aiming side. And the number of reflections at the same spatial position of the spots, that is, the number of overlapping spots, is denoted the reuse times of spot spatial position.
The spherical mirror on the incident side is a rectangular field mirror M3, and the aiming side is composed of two circular objective mirrors M1 and M2.
The distance between the incident side and the aiming side is equal to the curvature radius of the spherical mirrors.
The curvature centers C1 and C2 of the objective mirrors M1 and M2 are on M3 and do not coincide.
The curvature center C3 of the field mirror M3 is located at the center position of M1 and M2.

The reflection law of the classical PBWC is as follows: as a confocal resonator, the MPC refocuses the image of the entrance aperture onto the field mirror until the beam leaves the MPC.
Two columns of spots are obtained on the field mirror M3, and the total number of spots is $2 \times n_{PBWC}$, where $n_{PBWC}$ is the number of rows of spots on M3, and the reuse times of each spot spatial position is 1.
In Fig.\ref{fig:1} (a), the first and second columns of spots on M3 are focused by M2 and M1, respectively.
The number of spots on the objective mirrors M1 and M2 is 1. The reuse times of spot spatial position on M1 are $n_{PBWC}$, and the reuse times of spot spatial position on M2 are $n_{PBWC} + 1$.
The number of the ray passes of PBWC
\begin{equation}
 N_{PBWC}=2 \times (2 \times n_{PBWC} + 1)
\end{equation}
According to the above reflection law, the spot distribution on M3 is uniform, and the utilization ratio of the mirror is high, but the spatial position of each spot is reflected only once; however, the spatial position of the spots on M1 and M2 are reused for a more number of times, but there is only one spot, and the utilization ratio of the mirrors is low.

In view of the above defects and shortcomings, in order to simultaneously improve the reuse times of spot spatial position on the field mirror M3 and the utilization ratio of the objective mirrors M1 and M2 in the classical PBWC, we propose a design strategy of the dual circulation mode based on two-sided field mirrors. As shown in Fig.\ref{fig:1} (c), the origin of the Cartesian coordinate system is located at the midpoint of the line connecting the geometric centers of the mirrors on both sides, and the direction of the z-axis is along this line.
We made three major improvements to the classical PBWC.
(a) Two square spherical mirrors M4 and M5 are added at the original exit and entrance positions of the PBWC.
(b) The objective mirrors M1 and M2 are changed from circular to rectangular, so that M2 and the newly added M4 and M5 form a new PBWC.
(c) The entrance and exit positions are adjusted from the field mirror M3 side to the objective mirror M2 side.
Based on the above improvements, five rectangular spherical mirrors form two independent PBWCs, and a novel MMC based on PBWC-PBWC is obtained.

In this design, M2 is the objective mirror of the original PBWC, and each spot on M2 has high reuse times of spot spatial position; at the same time, M2 is also the field mirror of the new PBWC. Two columns of spots will be formed on M2, and two columns of mirror spots will be formed on M1, which greatly improves the utilization ratio of the mirror M1 and M2.
Using the principle of optical path reversibility, the new added objective mirrors M4 and M5 are set at the exit and entrance positions of the original PBWC, and the mutual nesting of the two PBWC reflection processes is realized. While maintaining high mirror utilization ratio of the M3, the reuse times of spot spatial position on the M3 are also effectively increased.
It can be seen that the design strategy of the dual circulation mode based on two-sided field mirrors realize the synchronous improvement of the utilization ratio of the mirror and the reuse times of spot spatial position on the two-sided spherical mirror under the premise of maintaining the simple structure of the MPC, thus improving the OPL.

Fig.\ref{fig:1} (c)(d) show the detailed configuration of the novel MMC based on PBWC-PBWC, and Cn represents the curvature center of the spherical mirror Mn.
The spatial positions of the curvature centers of the five spherical mirrors are set in such a manner that the C1 is at the geometric center of M3 on the aiming side, and the C2 is set directly below C1.
The C3 is located at the center of the boundary between the M1 and M2 on the incident side. The C4 is located at the geometric center of the M2 on the incident side, and C4 is on the same horizontal line as C3. The C5 is directly above C4.
The ray is incident from the position "in" above the M2 on the incident side, and the aiming point position "aim" is located at the geometric center of the M5 on the aiming side. Two columns of spots are formed on the M3, M4 and M5 on the aiming side. Four columns of spots are formed on the M1 and M2 on the incident side. The ray exits at the position "out" horizontally adjacent to the incident position "in" above M2.

\subsection{The law of ray reflection}
\begin{figure}[htbp]
\centering\includegraphics[width=9cm]{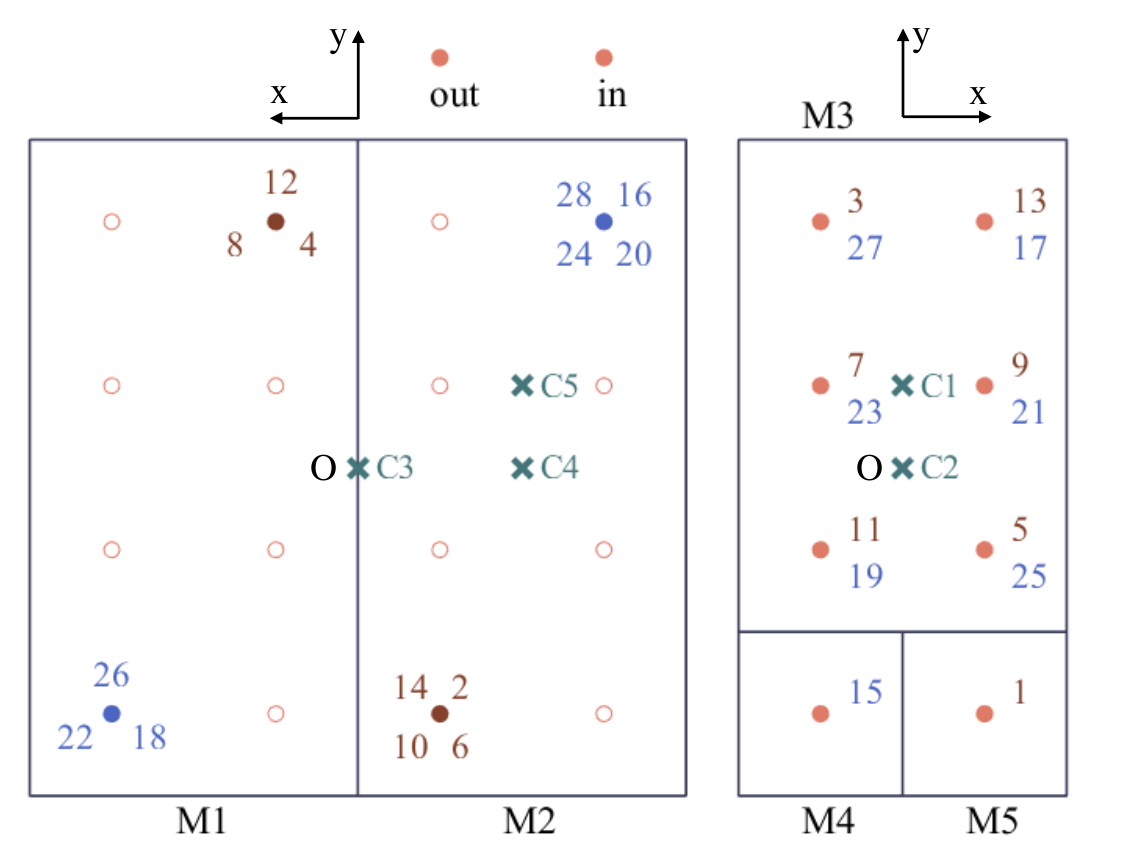}
\caption{First 28 spots of 4 rows-4 rows MMC based on PBWC-PBWC. The numbers represent the order of reflection. The numbers of the same color belong to the reflection of one PBWC-b.}
\label{fig:2}
\end{figure}

The five rectangular spherical mirrors of the MMC based on PBWC-PBWC form two independent PBWCs. For the convenience, this paper takes the MMC with 114 passes, in which the number of spot rows on both the incident side and the aiming side is 4, as an example to illustrate the law of ray propagation. The spot patterns on both sides are shown in Fig.\ref{fig:2}, and the numbers in the figure represent the order of spot formation.
We refer to the PBWC based on the field mirror M2 and the objective mirrors M4, M5 as PBWC-a, and the PBWC based on the field mirror M3 and the objective mirrors M1, M2 as PBWC-b.

\begin{figure}[htbp]
\centering\includegraphics[width=12cm]{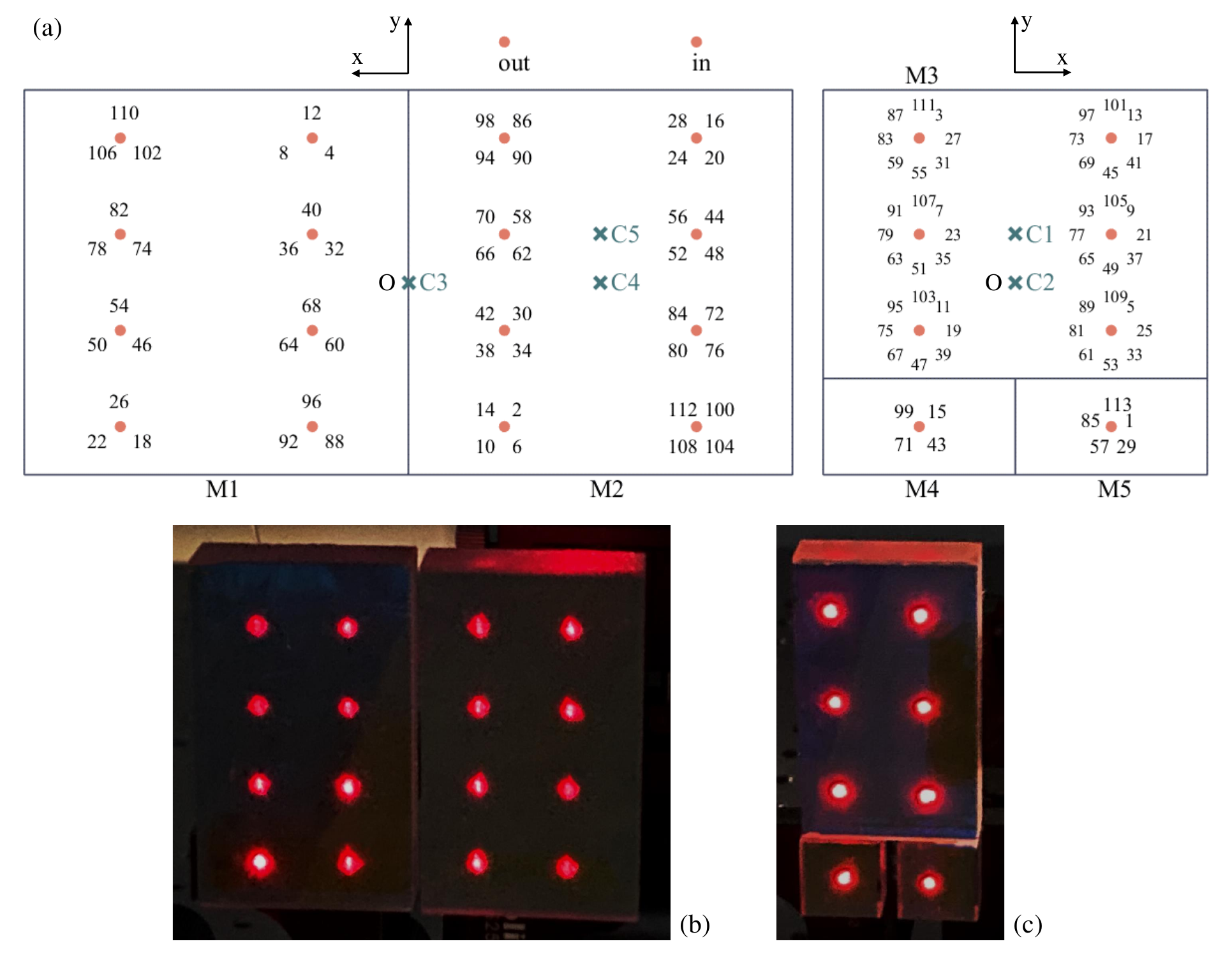}
\caption{Spot patterns of 4 rows-4 rows MMC based on PBWC-PBWC. (a) Reflection sequence of spots; (b) observed patterns on mirrors of the incident side; (c) observed patterns on mirrors of the aiming side.}
\label{fig:3}
\end{figure}

Fig.\ref{fig:3} (a) gives a detailed description of the complete spot formation sequence of the MMC with 114 passes.
Since M1 and M2 are the objective mirrors of the PBWC-b, they have the same number of 8 spot positions.
We build a 4 rows-4 rows MMC based on PBWC-PBWC using five rectangular spherical mirrors with curvature radius of 1m.
The laser is placed at the incident position of the MMC, and the observed spot patterns are shown in Fig.\ref{fig:3} (b)(c).

The propagation mode of the ray in the MMC is as follows: the ray enters the MMC from the incident position "in" of the PBWC-a, and the first reflection occurs at the position "1" which is located at the center of the objective mirror M5 of the PBWC-a and taken as the aiming point. The spot "2" is formed at a new position on the PBWC-a field mirror M2. The spot "2" and the incident point "in" are symmetrical about the curvature center C5 of M5. It can be understood as the first and second reflections of PBWC-a.
Since the PBWC-a objective mirror M5 is just set at the incident position of PBWC-b, the process of the ray from the spot "1" on M5 to the spot "2" on M2 is equivalent to the incident process of PBWC-b.

Next, a complete reflection process of the classical PBWC will be performed: starting from the initial spot "1" of the PBWC-b, on the field mirror M3 of the PBWC-b at the aiming side, a sequence of spots "1-3-5-7-11-13" will be formed in turn.
The spots "2-6-10-14" are formed on the objective mirror M2 of the PBWC-b at the incident side, and the reuse times of spot spatial position on M2 is 4. The corresponding spots are the spots "4-8-12" on the objective mirror M1 of the PBWC-b, and the reuse times of spot spatial position on M1 is 3.
That is, the 1st to 14th spots are the complete reflection process of the first PBWC-b.

Since the PBWC-a objective mirror M4 is just set at the exit position of PBWC-b, when the ray continues to shoot from the spot "14" on M2 to the spot "15" on M4, a new spot position "16" will be obtained on the field mirror M2 of the PBWC-a, which is symmetric with the spot "14" about the curvature center C4 of M4.
It can be understood as the third step reflection of PBWC-a.
Meanwhile, according to the principle of optical path reversibility, the process of the ray from the spot "15" on M4 to the spot "16" on M2 is equivalent to the re-incident of PBWC-b. 

Then the complete reflection process of the second PBWC-b will begin:
the spots "15-17-19-21-23-25" is sequentially formed on the field mirror M3 of the PBWC-b at the aiming side from the initial spot "15" of the PBWC-b.
The spots "16-20-24-28" are formed on the objective mirror M2 of the PBWC-b at the incident side, and the corresponding spots "18-22-26" are formed on the objective mirror M1 of the PBWC-b.
Then, the 15th to 28th spots are the complete reflection process of the second PBWC-b.

Therefore, we can get the law of light reflection in the MMC designed by the dual circulation mode based on two-sided field mirrors. It can be found that after each complete reflection of the PBWC-b, one step reflection of the PBWC-a is performed, and a new spot position is added on the field mirror M2 of the PBWC-a. This step of reflection is also the first step of reflection of the next PBWC-b. By repeating this process, after the reflection of the 8th PBWC-b is completed, the 114th spot and the 112th spot on M2 are symmetric about the curvature center C5 of M5, that is, the ray reaches the exit position "out" of the PBWC-a.

\subsection{Simple design method of MMC based on PBWC-PBWC}

According to the design strategy of the dual circulation mode based on two-sided field mirrors, the MMC based on PBWC-PBWC contains two classical PBWCs at the same time. The first cell in its name is the big cycle "a"cell, PBWC-a, and the second cell is the small cycle "b" cell, PBWC-b.
By changing the row number of spot pattern on the field mirrors of the PBWC-a and PBWC-b, the matrix spot pattern with arbitrary number of rows and different OPL can be realized.
Based on the basic principle of classical PBWC, a simple design method of MMC based on PBWC-PBWC is given, and the spot distribution and the mathematical formula of the total OPL of the MMC are derived.

More generally, if the total row number of spot pattern is $n_2$ on M3, M4 and M5 on the aiming side, then the 1st to $2 \times  (2 \times n_2 - 1 ) $ spots are the complete reflection process of the first PBWC-b.
The  $2 \times ( 2 \times n_2 -1  ) +1 $ to $4 \times ( 2 \times n_2 -1  ) $ spots are the complete reflection process of the second PBWC-b.
By repeating this process, if the row number of spot pattern is $n_1$ on M1 and M2 on the incident side, after the complete reflection of the $2\times n_1$ PBWC-b is completed, the  $2 \times ( 2 \times n_2 -1  ) \times 2 \times n_1 +2$ spot is located at the exit position "out" of the PBWC-a.

Since M2 is the field mirror of PBWC-a, the number of spots on M2 is equal to the number of complete PBWC-b.
Therefore, the reuse times of each spot spatial position on the field mirror M3 of the PBWC-b is equal to the number of complete PBWC-b , that is, $2 \times n_1$.
Based on the theory of the classical PBWC, the reuse times of spot spatial position on M1 and M2 are equal to $n_2-1$ and $n_2$, respectively, and the reuse times of spot spatial position on M4 and M5 are equal to $n_1$ and $n_1 + 1$, respectively.
Then, the number of ray passes of the MMC based on PBWC-PBWC is equal to the product of the number of ray passes of PBWC-b and the number of PBWC-b (that is, the number of spots on the field mirror of PBWC-a) plus 2.
That is, the number of ray passes
\begin{equation}
N_{PP}=[2\times(2\times n_2-1)] \times (2\times n_1)+2
\end{equation}
where $n_1$ and $n_2$ are both positive integers. Then, the OPL of the MMC
\begin{equation}
{\rm OPL}_{PP}=N_{PP}\times d
\end{equation}
where $d$ is the distance between the mirrors on both sides.

On the aiming side, the distance between the curvature centers C1 and C2 is half of the row spacing $d_{r2}$ of spots. By adjusting the positions of C1 and C2, the number of rows of the spots on the aiming side can be increased or decreased.
The distance between the aiming point "aim" and C1C2 in the $x$ direction is half of the column spacing $d_{c2}$ of spots.
On the incident side, the distance between the curvature centers C4 and C5 is half of the row spacing $d_{r1}$ of spots. By adjusting the positions of C4 and C5, the number of rows of the spots on the incident side can be increased or decreased.
The distance between the incident point "in" and C4C5 in the $x$ direction is half of the column spacing $d_{r1}$ of spots.
The distance between C3 and C4C5 in the $x$ direction is equal to the column spacing $d_{c1}$.

\begin{figure}[htbp]
\centering\includegraphics[width=14cm]{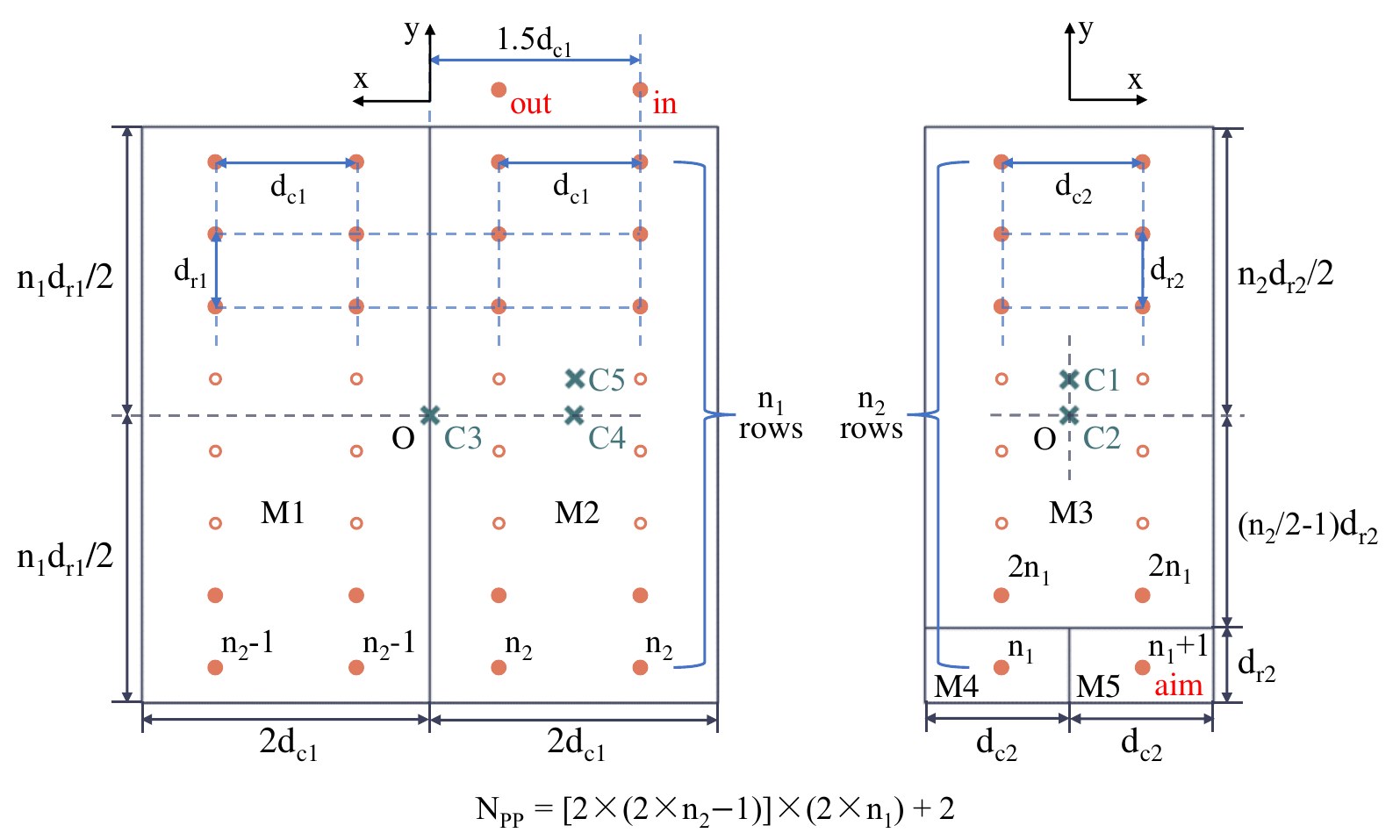}
\caption{ Design method of MMC based on PBWC-PBWC. The number marked on the upper right corner of the spot represents the reuse times of the spot spatial position.}
\label{fig:4}
\end{figure}

On the basis of the above reflection law, we propose a design method of the MMC based on PBWC-PBWC.
We denote the number of spots on each spherical mirror as $n_{spots}$, the reuse times of spot spatial position as $n_{re}$, and the distance between the mirrors on both sides as $d$.
The mirror coordinate axis and the parameters are shown in Fig.\ref{fig:4} and Table.\ref{tab:1}.
Then, the curvature center coordinates of the five spherical mirrors are: C1, C2 on the aiming side, C3, C4, C5 on the incident side.
The incident position "in" is above M2, and the aiming position "aim" is at the geometric center of M5.
Based on this design theory, MMCs with different number of spot rows and different spot spacing can be obtained, that is, MMCs with different OPL.
When $d=1m$, $n_1=20$, $n_2=20$, the MMC with three kilometers OPL can be obtained.

\begin{table}[htbp]
  \centering
  \caption{Design parameters of MMC based on PBWC-PBWC}
\resizebox{.99\columnwidth}{!}{
    \begin{tabular}{|c|c|c|c|c|c}
\hline
          & \multicolumn{1}{c|}{M1} & \multicolumn{1}{c|}{M2} & \multicolumn{1}{c|}{M3} & \multicolumn{1}{c|}{M4} & \multicolumn{1}{c|}{M5} \\
\hline
    Curvature  center & \multicolumn{1}{c|}{C1(0,$d_{r2}$/2,$-d$/2)} & \multicolumn{1}{c|}{C2(0,0,$-d$/2)} & \multicolumn{1}{c|}{C3(0,0,$d$/2)} & \multicolumn{1}{c|}{C4($-d_{c1}$,0,$d$/2)} & \multicolumn{1}{c|}{C5($-d_{c1},d_{r1}$/2,$d/2$)} \\
\hline
    $n_{spots}$     & \multicolumn{1}{c|}{2 $\times$ $n_1$} & \multicolumn{1}{c|}{2 $\times$ $n_1$} & \multicolumn{1}{c|}{2 $\times$ $n_2-2$} & \multicolumn{1}{c|}{1} & \multicolumn{1}{c|}{1} \\
\hline
    $n_{re}$     & \multicolumn{1}{c|}{$n_2-1$} & \multicolumn{1}{c|}{$n_2$} & \multicolumn{1}{c|}{2 $\times$ $n_1$} & \multicolumn{1}{c|}{$n_1$} & \multicolumn{1}{c|}{$n_1+1 $} \\
\hline
     Incident  position "in" & \multicolumn{5}{c|}{($-d_{c1} \times 1.5, d_{r1} \times (n_1+1)/2, d/2)$} \\
\hline
    Aiming position "aim" & \multicolumn{5}{c|}{($d_{c2}/2, -d_{r2} \times (n_2-1)/2, -d/2)$} \\
\hline
    Exit position "out"& \multicolumn{5}{c|}{$(-d_{c1}/2, d_{r1} \times (n_1+1)/2, d/2)$} \\
\hline
     The number of ray passes $N_{PP}$ & \multicolumn{5}{c|}{$[2\times (2\times n_2-1)]\times (2\times n_1)+2$} \\
\hline
    OPL   & \multicolumn{5}{c|}{$N_{PP}\times d $} \\
\hline
    \end{tabular}%
}
  \label{tab:1}%
\end{table}%

The optical stability of MPC is the main factor that restricts its performance and long-term reliability in actual measurement.
In practical applications, the interference of the external environment may have an adverse effect on the optical performance and stability of the MPC. For example, the vibration of the production site and other factors may cause a slight change in the direction of the incident ray, and the local deformation of MPC caused by the temperature change of the natural environment causes the deviation of the curvature center of the spherical mirror.
This paper will focus on the analysis of the angle sensitivity of the incident ray, the spot deformation and the position deviation of the curvature center of the spherical mirror.

The MMC based on PBWC-PBWC designed in this paper is a typical confocal cavity structure, which has two distinct advantages:
\begin{enumerate}
\item The exit position is not sensitive to the angle of the incident ray, when the paraxial approximation is satisfied. When the paraxial approximation is strictly satisfied, the position of the n th spot on the mirror can be obtained by the symmetry of the position of the (n-2) th spot with respect to the curvature center of the mirror where the (n-1) th spot is located. The aiming point "aim" is located on the spherical mirror M5. When the angle of the incident ray is offset, the position of the aiming point will be offset, and the spots on the aiming side mirrors will also be offset. As long as the position of each spot on the aiming side does not change, the formation of the spots on the incident side will not change, and the exit position will not change. Therefore, it is insensitive to the slight disturbance of the incident ray direction caused by environmental factors.
\item The beam quality is good and the spot deformation is small. The long OPL MMC designed in this paper is a confocal cavity. It using spherical mirrors with curvature radius of 1000 mm strictly satisfies the paraxial approximation condition. For the MMC with confocal structure, only two beam shapes appear alternately on the mirrors\cite{ozharar2017mirrors}. That is, there is one beam shape on the incident side mirrors and the other beam shape on the aiming side mirrors. The shape and size of the spots on both sides can be adjusted by adjusting the collimation degree of the incident ray. Therefore, the MMC has the advantages of small spot deformation and good beam quality. 
\end{enumerate}

If the MMC based on PBWC-PBWC proposed in this paper is designed to be miniaturized or portable, the aberration and spot deformation caused by the deviation from the paraxial approximation condition must be considered, and their impact on the measurement must be analyzed.

The position deviation of the curvature centers is an important factor affecting the performance of the MMC based on PBWC-PBWC. Since the five spherical mirrors have different functions in the PBWC, they have different sensitivities to environmental disturbances such as temperature and vibration.
\begin{itemize}
\item The spherical mirrors M1 and M2 are the objective mirrors of the PBWC-b, and their curvature centers C1 and C2 are located on the field mirror M3 of the PBWC-b. The C1 and C2 are used to form two columns of spots of the classical PBWC on the aiming side. The distance between C1 and C2 determines the row spacing of the spots on the aiming side.
In practical applications, the adjusted spherical mirrors M1 and M2 can be fixed as a whole, so as to ensure that the spot row spacing is unchanged.
The overall shift or deflection of the spatial positions of the C1 and C2 in the x and y directions will lead to the synchronous shift or deflection of the two columns of spots on the aiming side.
In the process of spot shifting or deflecting, if there is no spot exceeding the range of the mirror size or hitting the interface of the spherical mirrors, it will not have a great impact on the exit position, and will not affect the use of the MMC. The M1 and M2 show excellent stability.
\item The spherical mirror M3 is the field mirror of the PBWC-b, and its curvature center C3 is located at the center of the objective mirrors M1 and M2 of the PBWC-b on the incident side. The role of M3 is to realize the mirror mapping of the spots on M1 to M2, and it shows the most excellent stability in the x and y directions. The shift of the spatial position of the C3 mainly affects the spatial position of the spots on M1. Under the condition that the spot does not exceed the range of the mirror size or hit the interface of the spherical mirrors, the position of the emergent ray is almost not affected by the initial disturbance.
\item The spherical mirrors M4 and M5 are the objective mirrors of the PBWC-a, and their curvature centers C4 and C5 are located on the field mirror M2 of the PBWC-a on the incident side. Because the exit spot "out" of the MMC is located at the exit position of the PBWC-a, the position shift or deflection of the C4 and C5 will directly lead to the shift or deflection of position of the exit spot, and the offset is positively correlated with the number of reflections.
The C4 and C5 are the most sensitive to the environmental perturbations in the x and y directions.
\end{itemize}

Therefore, in order to reduce the influence of the position deviation of the curvature center on the MMC, we should fix the incident side M1 and M2 as a whole, and the aiming side M3, M4 and M5 as a whole, to ensure the synchronous change of the curvature centers of both sides, and adopting an optical frame with excellent structural stability and good adjustability will improve the stability and reliability of the MMC in practical applications.

\section{Extended design for MMCs based on multi-cycle mode of two-sided field mirrors}

Due to the design strategy of the dual circulation mode of two-sided field mirrors, considering the classical PBWC and BHWC, this paper proposes the MMCs based on PBWC-BHWC, BHWC-PBWC and BHWC-BHWC. Their simple design methods and optical configurations are given, and the calculation formula of the number of ray passes is summarized.
Furthermore, the design strategy proposed in this paper can be applied to three cycles or more, and a new general design method for ultra-long OPL MMCs is proposed.
The design of MMC based on PBWC-PBWC-PBWC with triple circulation mode is described in detail, which enriches the design theory of two-sided field mirrors.

\subsection{The MMCs based on PBWC-BHWC, BHWC-PBWC and BHWC-BHWC}

\begin{figure}[htbp]
\centering\includegraphics[width=8cm]{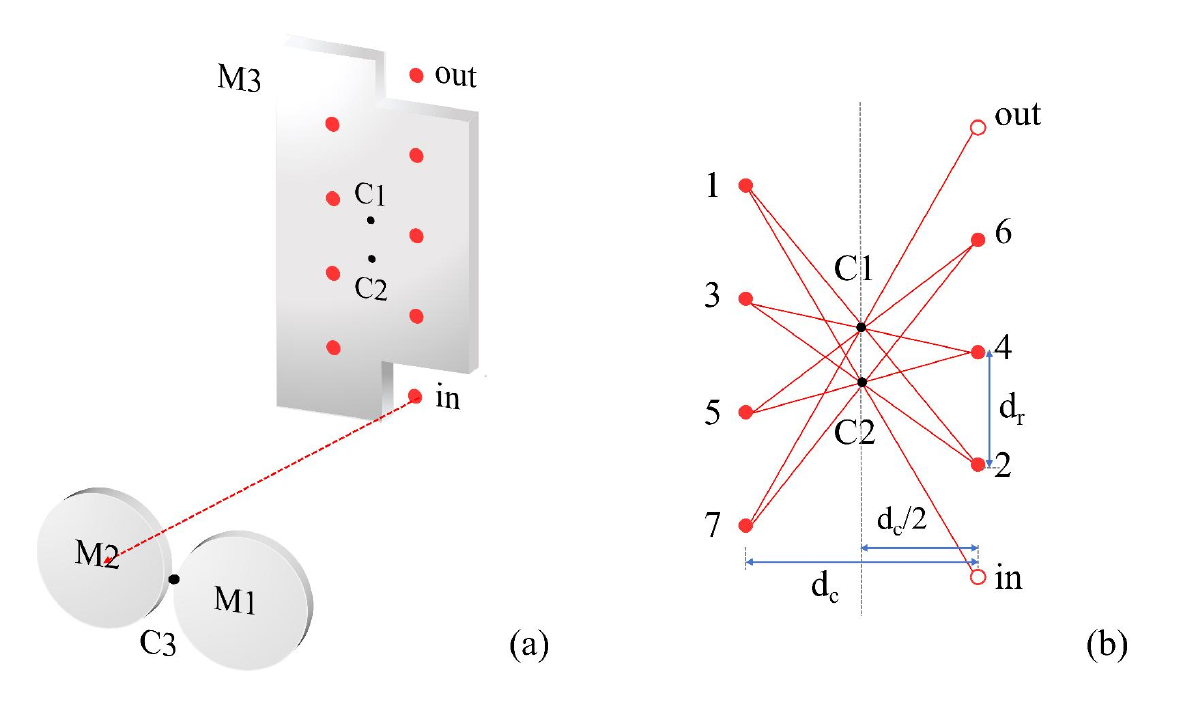}
\caption{ Configuration and spot pattern of the classical BHWC ($n_{BHWC} = 4$). (a) Configuration of the classical BHWC; (b) spot pattern on the field mirror M3 of the classical BHWC.}
\label{fig:5}
\end{figure}

The configuration and spot pattern of the classical BHWC are shown in Fig.\ref{fig:5}. The spherical mirror on the incident side of the classical BHWC is a rectangular field mirror M3 with two notches. The two notches correspond to the incident position and the exit position, respectively. The aiming side is composed of two circular objective mirrors M1 and M2.
Its reflection law is basically consistent with the classical PBWC, and two columns of cross-arranged spots are obtained on the field mirror M3, showing a trapezoidal distribution.
The total number of spots is $2\times n_{BHWC}-1$, where $n_{BHWC}$ is the number of spots on the short base of the trapezoid, and the reuse time of each spot spatial position is 1.
In Fig.\ref{fig:5}, the first and second columns of spots on the field mirror M3 are focused by M2 and M1, respectively.
The number of spots on the objective mirrors M1 and M2 is 1, and the reuse times of spot spatial position on M1 and M2 are $n_{BHWC}$.
The number of the ray passes of BHWC 
\begin{equation}
N_{BHWC}=2\times (2\times n_{BHWC})
\end{equation}

The utilization ratio of each spherical mirror and the reuse times of spot spatial position of the BHWC have the same defects as the PBWC.
Therefore, the BHWC is also suitable for the design strategy of the dual circulation mode based on two-sided field mirrors.
Different from the adjacent incident position and exit position of the PBWC, the incident position and exit position of the BHWC are located at both ends of the field mirror, which can provide more installation space for the laser transmitting and receiving device of TDLAS technology.

On account of the design concept of the dual circulation mode based on two-sided field mirrors, combining PBWC and BHWC, we propose MMCs based on PBWC-BHWC, BHWC-PBWC and BHWC-BHWC.
Their naming principles are the same as the aforementioned MMC. The first cell refers to the large circulating "a" cell, and the second refers to the small circulating "b" cell.
They all use five rectangular spherical mirrors, and the number of spherical mirrors on the incident side and the aiming side are 2 and 3, respectively. The distance between the mirrors on both sides is equal to the radius of curvature of the spherical mirror.
The reflection law of these three MMCs is the same as that of MMC based on PBWC-PBWC.
Two spherical mirrors are placed at the exit and entrance positions of the "b" cell, and they form "a" cell with one of the objective mirrors of the "b" cell.
As a result, every time the reflection of a small cycle "b" cell is completed, one step of a large cycle "a" cell is reflected, and this step of reflection is also the first step of reflection of the next small cycle "b" cell.

\begin{figure}[htbp]
\centering\includegraphics[width=11cm]{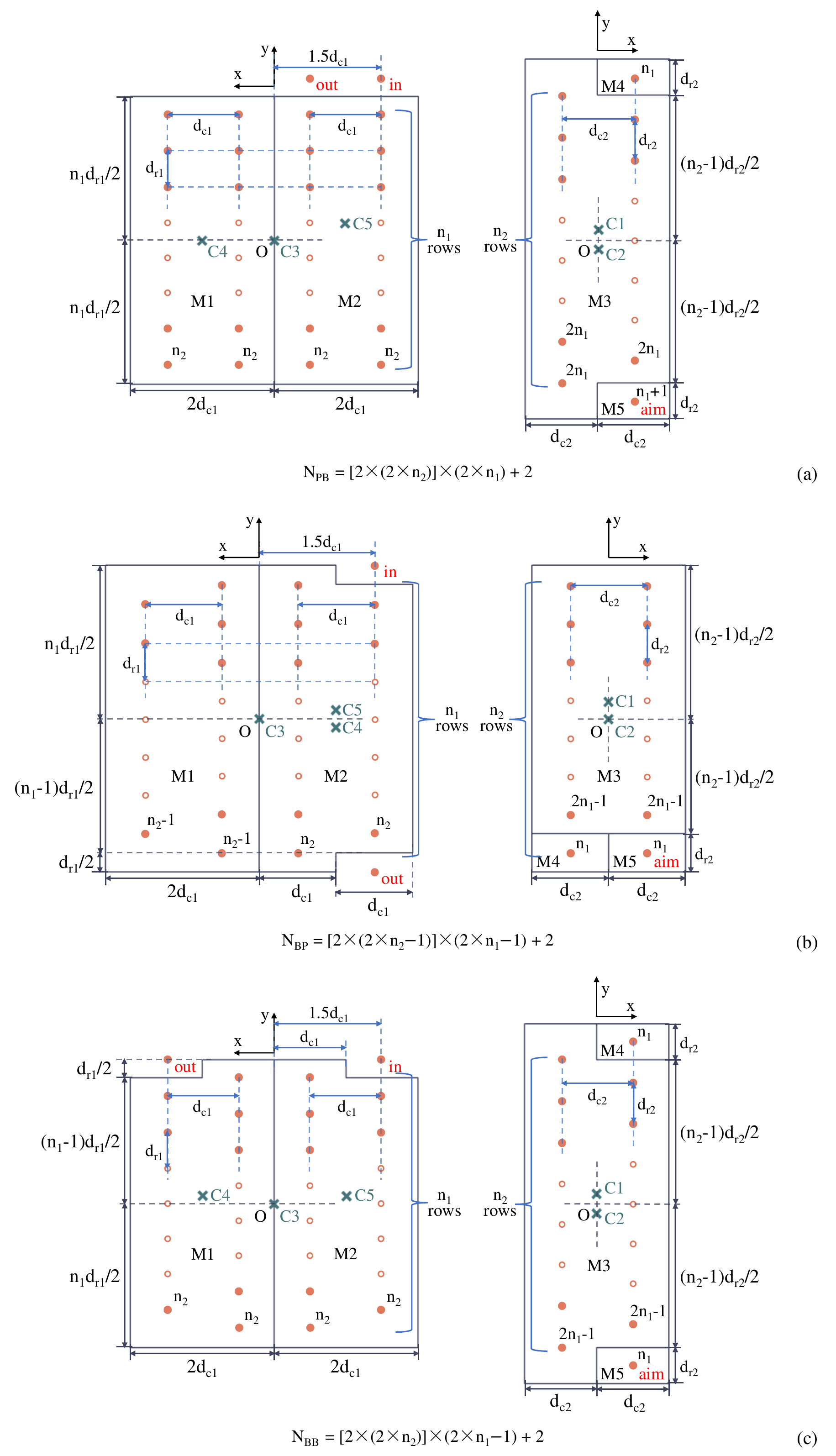}
\caption{ Design of MMCs based on PBWC-BHWC, BHWC-PBWC and BHWC-BHWC. The number marked on the upper right corner of the spot represents the reuse times of the spot spatial position. (a) MMC based on PBWC-BHWC; (b) MMC based on BHWC-PBWC; (c) MMC based on BHWC-BHWC.}
\label{fig:6}
\end{figure}

The design methods of the three kinds of MMCs are given.
We denote the number of rows of spots on the incident side as $n_1$, the row spacing of spots as $d_{r1}$, and make the column spacing of spots equal to $d_{c1}$; we denote the number of rows of spots on the aiming side as $n_2$, the row spacing of spots as $d_{r2}$, and the column spacing of spots as $d_{c2}$.
The mirror coordinate axes and parameters of the three MMCs are shown in Fig.\ref{fig:6} and Table.\ref{tab:2}.

\begin{table}[htbp]
  \centering
  \caption{Design parameters of MMCs based on PBWC-BHWC, BHWC-PBWC and BHWC-BHWC}
\resizebox{.99\columnwidth}{!}{
\renewcommand{\arraystretch}{2}
   \begin{tabular}{|c|c|c|c|c|c|c|}
         
\hline
    parameters & \multicolumn{2}{c|}{MMC based on PBWC-BHWC} & \multicolumn{2}{c|}{MMC based on BHWC-PBWC} & \multicolumn{2}{c|}{MMC based on BHWC-BHWC} \\
\hline
    a cell & PBWC  & M2-M4M5 & BHWC  & M2-M4M5 & BHWC  & M2-M4M5 \\
\hline
    b cell & BHWC  & M3-M1M2 & PBWC  & M3-M1M2 & BHWC  & M3-M1M2 \\
\hline
    M1    & $n_{spots}$ & $2\times n_1$  & $n_{spots}$ & $2\times n_1-1$ & $n_{spots}$ & $2\times n_1-1$ \\
          & $n_{re}$   & $n_2$    & $n_{re}$  & $n_2-1$  & $n_{re}$  & $n_2$ \\
\hline
    M2    & $n_{spots}$ & $2\times n_1$  & $n_{spots}$ & $2\times n_1-1$ & $n_{spots}$ & $2\times n_1-1$ \\
          & $n_{re}$   & $n_2$    & $n_{re}$   & $n_2$    & $n_{re}$   & $n_2$ \\
\hline
    M3    & $n_{spots}$ & $2\times n_2-1$ & $n_{spots}$ & $2\times n_2-2$ & $n_{spots}$ & $2\times n_2-1$ \\
          & $n_{re}$   & $2\times n_1$ & $n_{re}$   & $2\times n_1-1$ & $n_{re}$   & $2\times n_1-1$ \\
\hline
    M4    & $n_{spots}$ & 1     & $n_{spots}$ & 1     & $n_{spots} $ & 1 \\
          & $n_{re}$   & $ n_1$    & $n_{re}$   & $ n_1$    & $n_{re}$   & $ n_1$ \\
\hline
    M5    & $n_{spots}$ & 1     & $n_{spots}$ & 1     & $n_{spots}$ & 1 \\
          & $n_{re}$   & $n_1+1$  & $n_{re}$   & $n_1$    & $n_{re}$   & $ n_1$ \\
\hline
    C1    & \multicolumn{2}{c|}{$(0, d_{r2}/4, -d/2)$} & \multicolumn{2}{c|}{$(0, d_{r2}/2, -d/2)$} & \multicolumn{2}{c|}{$(0, d_{r2}/4, -d/2)$} \\
\hline
    C2    & \multicolumn{2}{c|}{$(0, -d_{r2}/4, -d/2$)} & \multicolumn{2}{c|}{$(0, 0, -d/2)$} & \multicolumn{2}{c|}{$(0, -d_{r2}/4, -d/2)$} \\
\hline
    C3    & \multicolumn{2}{c|}{$(0, 0, d/2)$} & \multicolumn{2}{c|}{$(0, 0, d/2)$} & \multicolumn{2}{c|}{$(0, 0, d/2)$} \\
\hline
    C4    & \multicolumn{2}{c|}{$(d_{c1}, 0, d/2$)} & \multicolumn{2}{c|}{$(-d_{c1}, -d_{r1}/4, d/2)$} & \multicolumn{2}{c|}{$(d_{c1}, d_{r1}/4, d/2)$} \\
\hline
    C5    & \multicolumn{2}{c|}{$(-d_{c1}, d_{r1}/2, d/2)$} & \multicolumn{2}{c|}{$(-d_{c1}, d_{r1}/4, d/2)$} & \multicolumn{2}{c|}{$(-d_{c1}, d_{r1}/4, d/2)$} \\
\hline
    Incident position "in" & \multicolumn{2}{c|}{$(-d_{c1}\times1.5, d_{r1}\times(n_1+1)/2, d/2)$} & \multicolumn{2}{c|}{$(-d_{c1}\times1.5, d_{r1}\times n_1/2, d/2)$} & \multicolumn{2}{c|}{$(-d_{c1}\times 1.5, d_{r1}\times n_1/2, d/2)$} \\
\hline
    Aiming position "aim" & \multicolumn{2}{c|}{$(d_{c2}/2, -d_{r2}\times n_2/2, -d/2)$} & \multicolumn{2}{c|}{$(d_{c2}/2, -d_{r2}\times (n_2-1)/2, -d/2)$} & \multicolumn{2}{c|}{$(d_{c2}/2, -d_{r2}\times n_2/2, -d/2)$} \\
\hline
    Exit position "out" & \multicolumn{2}{c|}{$(-d_{c1}/2, d_{r1}\times(n_1+1)/2, d/2)$} & \multicolumn{2}{c|}{$(-d_{c1}\times1.5, -d_{r1}\times n_1/2, d/2)$} & \multicolumn{2}{c|}{$(d_{c1}\times1.5, d_{r1}\times n_1/2, d/2)$} \\
\hline
    The number of ray passes $N$ & \multicolumn{2}{c|}{$N_{PB} =[2\times(2\times n_2)]\times (2\times n_1)+2$} & \multicolumn{2}{c|}{$N_{BP} =[2\times(2\times n_2-1)]\times(2\times n_1-1)+2$} & \multicolumn{2}{c|}{$N_{BB} =[2\times(2\times n_2)]\times(2\times n_1-1)+2$} \\
\hline
    OPL   & \multicolumn{2}{c|}{$N_{PB} \times d$} & \multicolumn{2}{c|}{$N_{BP} \times d$} & \multicolumn{2}{c|}{$N_{BB} \times d$} \\
\hline
    \end{tabular}%
}
  \label{tab:2}%
\end{table}%

\subsection{General design method for long OPL MMCs based on multi-cycle mode}

Based on the design strategy of two-sided field mirrors, the simple design method and optical configuration of four MMCs are given. The above four MMCs are composed of two cycles.
It is worth emphasizing that the design strategy proposed in this paper can be further extended to three cycles or more.
Taking the MMC based on PBWC-PBWC as an example, this paper describes how to expand it to a three-cycle MMC based on PBWC-PBWC-PBWC.

\begin{figure}[htbp]
\centering\includegraphics[width=12cm]{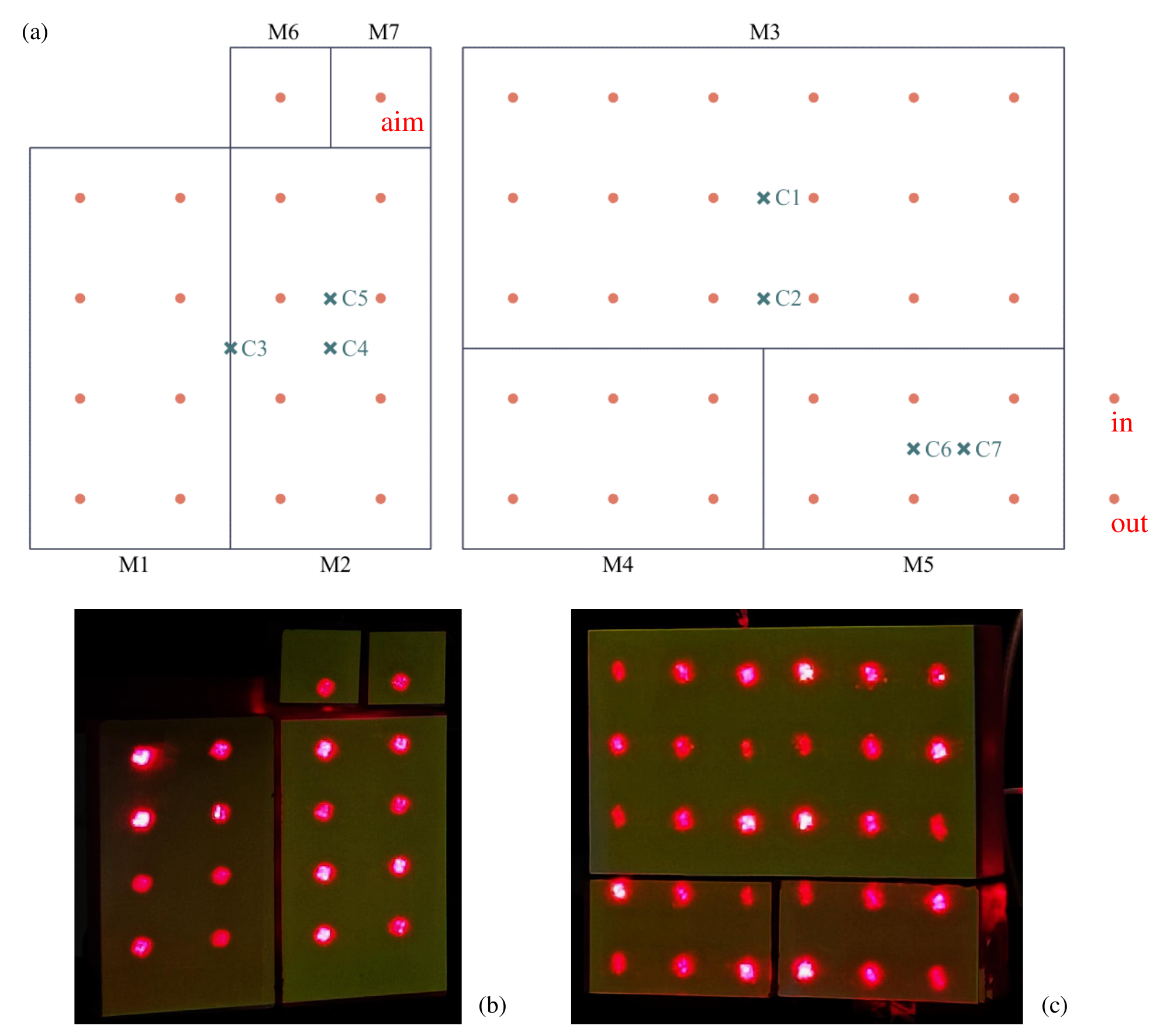}
\caption{Simulated and observed patterns of the MMC based on PBWC-PBWC-PBWC with 398 passes. (a) Spot pattern obtained by MATLAB simulation; (b) observed spot pattern on mirrors on the aiming side; (c) observed spot pattern on mirrors on the incident side.}
\label{fig:7}
\end{figure}

The general design method are as follow: 
Two new rectangular spherical mirrors M6 and M7 are placed at the exit position "out" and the incident position "in" of the original MMC based on PBWC-PBWC. The M6 and M7 are used as the two objective mirrors of the third cycle PBWC-c, and the rectangular spherical mirror M5 where the aiming point of the original MMC is used as the field mirror of the third cycle PBWC-c. The curvature center C6 of M6 is adjusted to the center position of M5. The curvature center C7 of M7 is adjusted to the right side of C6 on M5. The incident ray is adjusted to the right side of M5, and the center position of M7 is used as the aiming point. Then the MMC based on PBWC-PBWC-PBWC is constructed.
Using the above general design method, the MMC based on PBWC-PBWC with 114 passes is extended to the MMC based on PBWC-PBWC-PBWC with 398 passes. Fig.\ref{fig:7} shows the simulated and observed spot patterns.

\begin{figure}[htbp]
\centering\includegraphics[width=13cm]{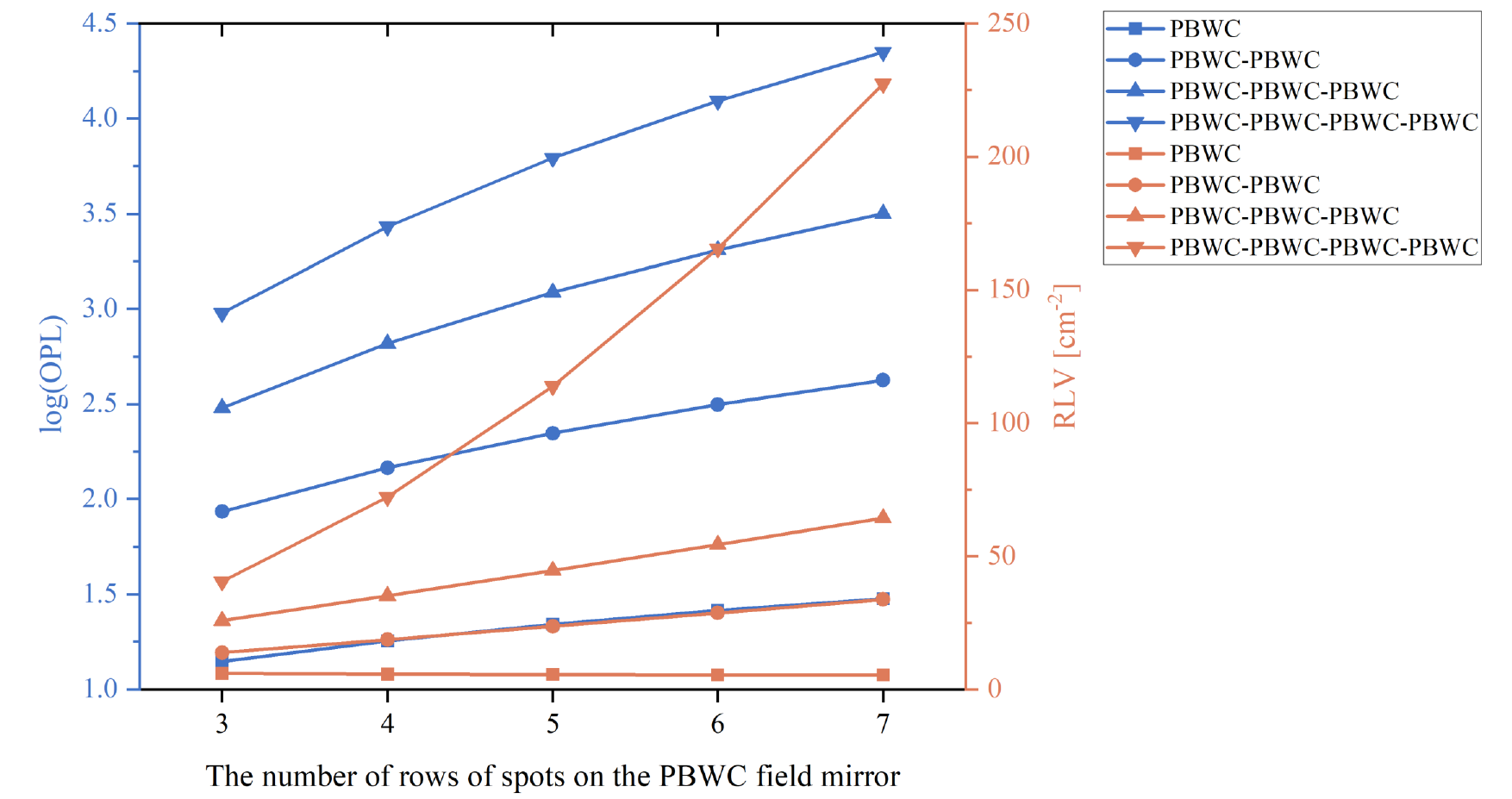}
\caption{ The OPL and RLV of the MMCs using the n PBWC circulation mode with different rows of the PBWC. The OPL of this figure is the OPL of the MMCs with mirror spacing d = 1m.}
\label{fig:8}
\end{figure}

We draw a line chart of the OPL and RLV of the MMCs using multi-PBWC circulation mode with different rows of the PBWC when the distance between mirrors is 1 m, as shown in Fig.\ref{fig:8}.
The RLV of the existing MPCs based on the two spherical mirrors with the independent circles or concentric circles spot patterns is in the range of 14.5-20\cite{ref32,ref33}, and the MMC using the triple PBWC circulation mode can achieve the same RLV.
The more the number of cycles, the better the effect of increasing the number of ray passes of the cyclic element on the increase of the OPL and RLV. When there are 7 rows of spots on the PBWC field mirror, the MMC using the quadruple PBWC circulation mode has an OPL of up to 22374 m and RLV is 227.29 cm$^{-2}$.
Therefore, the design strategy of multi-cycle mode has great potential for improving RLV of the MMC.
Increasing the number of cycles and the number of ray passes of the cyclic element significantly improves the OPL and RLV of the MMC.

The design method proposed in this paper provides a new idea for the design of MPC. For the single-sided field mirror MMC with incident position and exit position on the same side but do not coincide, such as Chernin multipass matrix systems (CMMS), the design strategy of multi-cycle mode can be used to increase the OPL of MMC from hundreds of meters to kilometers or even tens of kilometers. The designed new MMC have great potential application value in the field of high-precision trace gas monitoring.

\section{ CH$_4$ detection by using the PBWC-PBWC-PBWC type MMC}
\begin{figure}[htbp]
\centering\includegraphics[width=12cm]{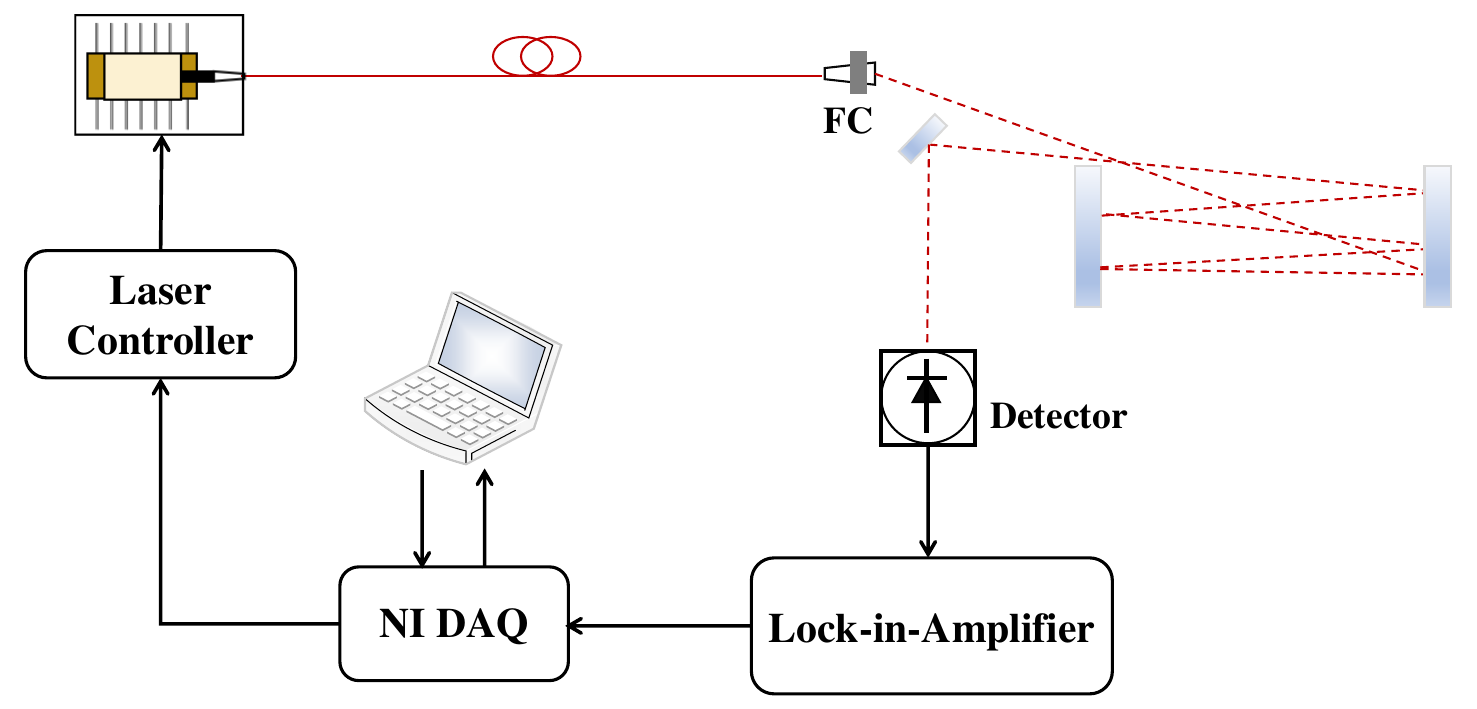}
\caption{ Schematic diagram of methane sensor (FC: Fiber Collimator).}
\label{fig:9}
\end{figure}

In order to verify the effectiveness of the design method of the novel ultra-long OPL MMC proposed in this paper, we use wavelength modulation spectroscopy (WMS) technology to detect trace methane gas in the laboratory ambient air with the MMC based on PBWC-PBWC-PBWC.
The experimental system is shown in Fig.\ref{fig:9}. We used a laser controller (ILX Lightwave, LDC-3724C) to control the driving current and temperature of a 1653 nm DFB laser (NLK1U5FAAA, NEL).
The methane absorption line of 6046.96 cm$^{-1}$ was selected as the target spectral line, and the current and temperature were set to 52.94 mA and 23.52 $^{\circ}$C, respectively.
By the LabVIEW program, a sawtooth wave signal with a frequency of 100 Hz and an amplitude of 1.5 V was generated, and a sine wave signal with a frequency of 8 kHz and an amplitude of 0.5 V is superimposed on the sawtooth wave. The laser drive current is scanned and modulated by a multi-functional DAQ card (NI-BNC-2110).
 
 In the experiment, the MMC based on PBWC-PBWC-PBWC used spherical mirrors with a radius of curvature of 1m, and the mirror reflectivity is 99.7 \%. The arrangement of mirrors and the spot pattern is shown in Fig.\ref{fig:7}, and the effective OPL is 398 m. The laser beam entered the MMC through the beam collimator. After 397 reflections, the ray was emitted from the position below the incident point, and the emergent ray was received by the InGaAs photodetector (Thorlabs, PDA50B2).
The original signal is collected by the host built-in data acquisition card, and the sampling rate is 1 MHz.
In the part of signal processing, the function of digital lock-in amplifier is realized by using LabVIEW program combined with DAQ card.

\begin{figure}[htbp]
\centering
\includegraphics[width=12cm]{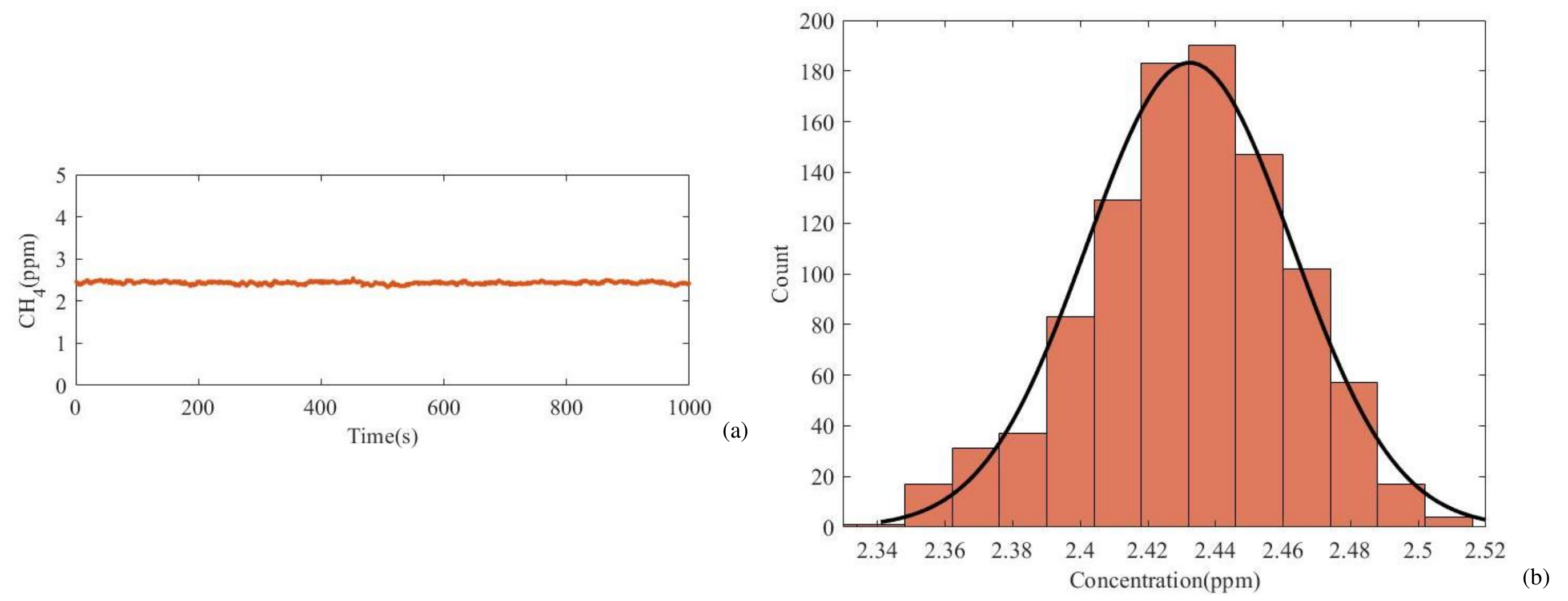}
\caption{(a) Time series measurements of methane in ambient laboratory air; (b) histogram plot for methane concentrations.}
\label{fig:10}
\end{figure}

To establish the relationship between the signal amplitude and the methane concentration, we calibrated the concentration with 2500 ppm·m methane standard gas (customized by Wavelength References Inc) at atmospheric pressure to analyze the 1000 data points obtained by the experiment.
The corresponding methane concentration obtained from the calibration is shown in Fig.\ref{fig:10} (a).
The methane concentration data show a Gaussian distribution with a 1$ \sigma $ standard deviation of 30.5 ppb, and the results are shown in Fig.\ref{fig:10} (b).
The experimental results verify the effectiveness of the design method of the MMC. This type of MMC has broad application prospects in the field of high-precision measurement of trace gases.

\section{Conclusions}

In this study, we first propose a general design method of ultra-long OPL MMCs based on multi-cycle mode of two-sided field mirrors.
Using the typical structure of PBWC and BHWC for reference, a design strategy of the dual circulation mode based on two-sided field mirrors is proposed, and the simple design methods of four kinds of MMCs based on dual circulation mode of PBWC and BHWC are given. Furthermore, a general design method for ultra-long OPL MMCs by adding cyclic elements is proposed, which greatly improves the mirror utilization ratio of the traditional MPC.
The OPL of the MMCs designed by this method can reach the order of kilometers or even tens of kilometers.
The novel MMCs have the advantages of simple structure, strong spot formation regularity, easy expansion, high mirror utilization ratio, high reuse times of spot spatial position, good stability and extremely high RLV.

A new MMC based on PBWC-PBWC-PBWC was established in the laboratory, and an open-path gas sensor was constructed based on this new MMC, which was used to continuously measure the methane in the laboratory ambient air, and the feasibility, effectiveness and practicability of the new design method were verified.
The design method proposed in this paper provides a new idea for the design of MPC, and the new designed have great potential application value in the field of high-precision trace gas monitoring, such as greenhouse gas monitoring, respiration diagnosis, semiconductor process and air pollution control.

\section*{Funding}
Beijing Define Technology Co., Ltd. (KJHX2018207, KJHX2020054, 2222000147)
\section*{Acknowledgments}
 This work was supported by Innovation Program for Beijing Define Technology Co., Ltd (Grant No. KJHX2018207, KJHX2020054, 2222000147)
\section*{Disclosures}
 The authors declare no conflicts of interest.
\section*{Data Availability Statement}
 Data underlying the results presented in this paper are not publicly available at this time but maybe obtained from the authors upon reasonable request.

\bibliography{sample}

\end{document}